# Tellurium substitution effect on superconductivity of the α-phase Iron Selenide

Kuo-Wei Yeh[1], Tzu-Wen Huang[1], Yi-Lin Huang[1], Ta-Kun Chen[1], Fong-Chi Hsu[1,2], Phillip M. Wu[3], Yong-Chi Lee[1], Yan-Yi Chu[1,2], Chi-Liang Chen[1], Jiu-Yong Luo[1], Der-Chung Yan[1] and Maw-Kuen Wu[1]

[1]*Institute of Physics, Academia Sinica, Nankang, Taipei, Taiwan*

[2]*Department of Materials Science and Engineering, National Tsing Hua University, Hsinchu, Taiwan*

[3]*Department of Physics, Duke University, Durham, NC, USA*

**Abstract**

We have carried out a systematic study of the PbO-type compound $FeSe_{1-x}Te_x$ ($x$ = 0~1), where Te substitution effect on superconductivity is investigated. It is found that superconducting transition temperature reaches a maximum of Tc=15.2K at about 50% Te substitution. The pressure-enhanced Tc of $FeSe_{0.5}Te_{0.5}$ is more than 10 times larger than that of FeSe. Interestingly, FeTe is no longer superconducting. A low temperature structural distortion changes FeTe from triclinic symmetry to orthorhombic symmetry. We believe that this structural change breaks the magnetic symmetry and suppresses superconductivity in FeTe.

**Introduction**

Recently, superconductivity was observed in α-FeSe$_{1-x}$ with critical temperature ~8K[1]. The crystal lattice of α-FeSe belongs to tetragonal (P4/*nmm*) symmetry at room temperature and is composed of a stack of edge-sharing FeSe$_4$-tetrahedra layer by layer [2,3]. The tetragonal phase α-FeSe with PbO-type structure has the same planar sublattice as the layered Fe based quaternary oxypnictides, which is the first system where Fe-element plays the key role to the occurrence of superconductivity [4]. Following the FeSe discovery, it is natural to ask whether chemical substitutions, either to the Se-site or the Fe-site, have any effects on superconductivity. Here we report the systematic study of the PbO-type compound FeSe$_{1-x}$Te$_x$ ($x$ = 0~1), where Te substitution effect on superconductivity is investigated. It is found that superconducting transition temperature increases with Te doping, reaches a maximum at about 50% substitution, and then decreases with more Te doping. Interestingly, FeTe is no longer superconducting. For FeSe, a structural transformation from tetragonal (P4/*nmm*) symmetry to triclinic (P-1) symmetry at around 105 K, which changes the lattice parameters without breaking magnetic symmetry, was reported and believed to have strong correlation with the occurrence of superconductivity. This Te substitutions study further confirms that the low temperature structural deformation is essential to the competition between magnetism and superconductivity in this class of materials.

**Experimental Methods**

Powder materials of Fe (3N purity), Se (3N purity) and Te (5N purity) with appropriate stoichiometry (FeSe$_x$Te$_{1-x}$ of $x$ = 0 ~ 1.0) were mixed in a ball mill for 1 h. The well-mixed powders were cold-pressed into discs under 400 kg/cm$^2$ uniaxial pressure, and then sealed in an evacuated quartz tube with a pressure less than 10$^{-4}$ torr and heat treated at 600 °C for 20 hours. The reacted bulk sample was reground into fine powders, re-pressed, sealed, and subsequently sintered at 650 ºC for at least 20 hours. Phase identification of the FeSe$_{1-x}$Te$_x$ powder samples was carried out on a diffractometer (Philip, PW3040/60) with Cu-Kα radiation generated at 45 kV and 40 mA. The *a*, *c* and gamma angle cell parameters were calculated from results of diffraction measurements using synchrotron source (BL12b2 at SPring 8) with incident beam of wavelength 0.995 Å. DC magnetic susceptibility measurements were performed in a Quantum Design superconducting quantum interference device vibrating sample magnetometer (SQUID VSM). The resistance measurements were carried out in a Quantum Design PPMS system, model 6000 using the standard 4-probe method with silver paste for contact.

**Result and Discussion**

FeSe with tetragonal PbO-type structure was reported to undergo a phase transformation toward a hexagonal-NiAs structure if synthesized at around 731 K [5]. In contrast, FeTe with the same tetragonal crystal structure is stable up to a much higher temperature, ~1200 K. One would then expect that the substitution of Se atoms within FeSe with Te will stabilize the tetragonal phase at a synthetic temperature close to or above 731 K. This agrees well with our X-ray diffraction analysis. Figure 1 shows powder diffraction patterns of the FeSe$_{1-x}$Te$_x$ series synthesized at 923 K (650 ºC). The inset shows the (101) diffraction peaks of various x concentration samples. For $x \geq 0.3$, diffraction peaks all belong to the tetragonal phase and these samples show good quality without second phases.  For low concentration of tellurium substitution, $x \leq 0.2$, the hexagonal-NiAs-type FeSe was present, with the strongest peak at $2\theta = 42.2°$. It is noteworthy that the sample $x = 0.2$ contains two tetragonal phases. One of the tetragonal phase which contains lower concentration of Te (not $\beta$ phase) can be eliminated by additional sintering in the $x = 0.3$ sample (all other samples were sintered twice only, see experimental section).  This phenomenon should be the result of the fact that Te, which has larger atomic size, inhibits interatomic diffusion in FeSe lattice.  In contrast, Se atoms move more easily in the larger FeTe lattice.  That is why the samples with higher Te concentration ($x \geq 0.5$) show only one tetragonal phase.

In Fig. 2 we present the temperature dependence of the normalized resistance for samples with varied Te concentrations at zero magnetic field. The inset displays the details near superconducting transition temperature. For x < 0.2, the samples show metallic behavior in the normal state. For x greater than 0.3 the normal state resistance increases with temperature decrease before the onset of superconductivity. The reason for this metal-semiconductor transition found in the $FeSe_{1-x}Te_x$ series is unclear at this stage. For x=0.2 sample, a sharp drop in resistance was observed at about 13.7 K, which indicates the onset of superconductivity. This value is higher than the onset transition temperature for FeSe. Tc is found to increase with $x$ and reaches a maximum of 15.2 K at $x = 0.5$.  In the meantime, a sharper superconducting transition is observed for samples with $x = 0.5$ and 0.7, which suggests these samples are more homogeneous. It is interesting to note that for these higher Te concentration samples, the normal state resistance is semiconductor-like even though the Tc increases. Tc decreases with further increase of Te-doping so that Tc of $FeSe_{0.1}Te_{0.9}$ drops to 11.4 K. It is surprising that FeTe is no longer superconducting even though its structure remains to be PbO-type. The magnetic susceptibility of $FeSe_{1-x}Te_x$ as a function of temperature was measured at 30 Gauss field. A sharp drop indicating the magnetic onset of superconductivity appears at ~8 K for FeSe, and ~12 K for $FeSe_{0.8}Te_{0.2}$.

The Tc variation as a function of pressure for FeSe and $FeSe_{50}Te_{50}$ samples has been determined from AC magnetic susceptibility measurements [6]. It is particularly interesting to note that the pressure-enhanced Tc of $FeSe_{0.5}Te_{0.5}$ ($dT_c/dP$ = 0.487 K/kbar) is more than 10 times larger than that of FeSe ($dT_c/dP$ = 0.0384 K/kbar) [6]. The pressure enhancement observed in FeSe is smaller than for results recently reported [7]; this is probably due to the difference in measurements. The results from Mizuguchi et al. [7] are derived from the variation in onset of resistive transition, whereas in our case the value is determined from the onset of magnetic transition. The $T_c$ enhancement by pressure most likely originates from the structural deformation observed at low temperature.

In order to examine how the Te-doping affects the low temperature structural change observed in FeSe, detailed X-ray refinements at room and low temperatures were carried out. As shown in Fig. 3(a), the lattice expands due to the fact that the ionic radius of Te is larger than that of Se. The expansions in $a$ and $c$ axis are normalized to the cell parameters of FeSe ($a$ = 3.775 Å and $c$ = 5.512 Å). It is clearly seen that for $FeSe_{0.5}Te_{0.5}$ the tetragonal lattice expands asymmetrically, with the $c$ axis expanding ~ 8.1%, and the $a$ axis expanding only about 0.6%. As Te concentration increases, the unit cell further expands, and the distance of Fe-Fe bond length in Fe-occupied plane extends. The increase in c-parameter may lead to an increase in the density of states at the Fermi level, and thus enhances Tc, as suggested by the recent density functional calculation [8]. As shown in Fig. 3(b), we observe that $T_c$ and γ angle of the distorted lattice both reach a maximum at the point of 50% Te doping. This correlation between the $\gamma$ angle and transition temperature $T_c$ leads us to believe that $T_c$ more critically depends on the level of lattice distortion than the distance of Fe-Fe bond in Fe-plane.

Figure 4 displays temperature dependence of the X-ray powder diffraction for FeSe$_{0.5}$Te$_{0.5}$. Data were collected using synchrotron source (BL12b2 at SPring 8) with incident beam of wavelength 0.995 Å. At room temperature the FeSe$_{0.5}$Te$_{0.5}$ crystal lattice already distorts to form the structure with space group P-1 as seen in Fig. 4(a). This is closely correlated with what is observed for the FeSe at temperatures below 105 K [1]. This result indicates that partially substituted Te-atom induces a local strain in the lattice. As the temperature is lowered to 6 K, unlike what was reported in LaO$_{0.9}$F$_{0.1}$FeAs [7], we do not observe the shrinkage of d-spacing along *a* and *b* direction (*a* = 3.809 Å, *c* = 5.995 Å and γ = 90.064º at 300 K; *a* = 3.801 Å, *c* = 5.950 Å and γ = 90.112º at 6 K, respectively). In addition, calculations from the diffraction data show the Te occupancy at Se sites was 53.58 %. This correlates well with our EDS chemical analysis results.

Figure 4(b) shows the Bragg peaks of $(220)$, $(203)$ and $(221)$ reflections of FeSe$_{0.5}$Te$_{0.5}$ measured at various temperatures. Two peaks belonging to (221) and $(2\bar{2}1)$ are distinctly separate for all temperatures, which indicates a deformation from P4/*nmm* to P-1 symmetry. Similar results of planar group (220) and $(2\bar{2}0)$ are also obvious in this graph. The (203) Bragg peaks did not split; therefore, this result strongly suggests that the deformation simply belongs to in-plane characteristic rather than out-of-plane behavior [9]. Moreover, the inset of Figure 4(b) reveals that the (211) Bragg peak begins to split at around 100 K (shown by an arrow). Therefore, the crystal distorts again to exhibit a larger gamma angle when the temperature is lower than 100 K, indicating a further distortion of P-1 lattice structure. These results strongly suggest that the structural distortion at low temperature is more important for understanding the origin of superconductivity. The much larger pressure effect on Tc in Te-doped samples can be possibly due to its relatively large c-axis, which allows for the further increase in gamma angle under pressure.

As previously reported, tetragonal FeSe compounds distort to a triclinic lattice below 105K with $a = b$ and $\gamma = 90.3°$. Surprisingly but interestingly, FeTe, which is tetragonal at room temperature, is not superconducting and exhibits magnetic order at about 60K. Detailed structural study at low temperature shows that FeTe transforms from tetragonal to orthorhombic with $a = 3.824$ Å; $b = 3.854$ Å; $c = 6.308$ Å and $\gamma = 90°$ at temperature below 45 K. The results indicate a different temperature dependence of lattice distortion between FeSe and FeTe. As shown schematically in Fig. 5(a), the lattice asymmetry in FeSe along the (110) axis, which distorts to a triclinic lattice at low temperature, does not destroy the magnetic symmetry. This suggests that magnetic frustration persists down to low temperature so that superconductivity is allowed to occur. On the other hand, as seen in Fig. 5(b), the lattice distortion of FeTe results in different lengths on *a* and *b* axis. This asymmetry along the (110) direction breaks the magnetic symmetry at low temperature so that a magnetic order sets in to prohibit the appearance of superconductivity. These results indicate that superconductivity is competing with a magnetic order in this class of material. Further studies using neutron scattering and structure information at high pressure to confirm this viewpoint are obviously urgently needed.

**Conclusion**

In summary, we find Te doping to the layered PbO-type α-FeSe modifies the superconductivity, with Tc at a maximum of ~15 K when Te replaces 50% of the Se. $T_c$ enhancement is well correlated with the structural deformation resulting from Te substitution. The presence of superconductivity is also closely associated with the magnetic symmetry of the Fe plane. The breaking of magnetic symmetry, such as that observed at low temperature in FeTe compound, results in the destruction of superconductivity. It will be interesting to investigate whether a superconducting transition can be observed in FeTe by suppressing the magnetic symmetry breaking, such as with an applied uniaxial pressure.

**Acknowledgements** We wish to thank the National Science Council of Taiwan and the US AFOSR through its Tokyo Office AOARD for their generous financial support.

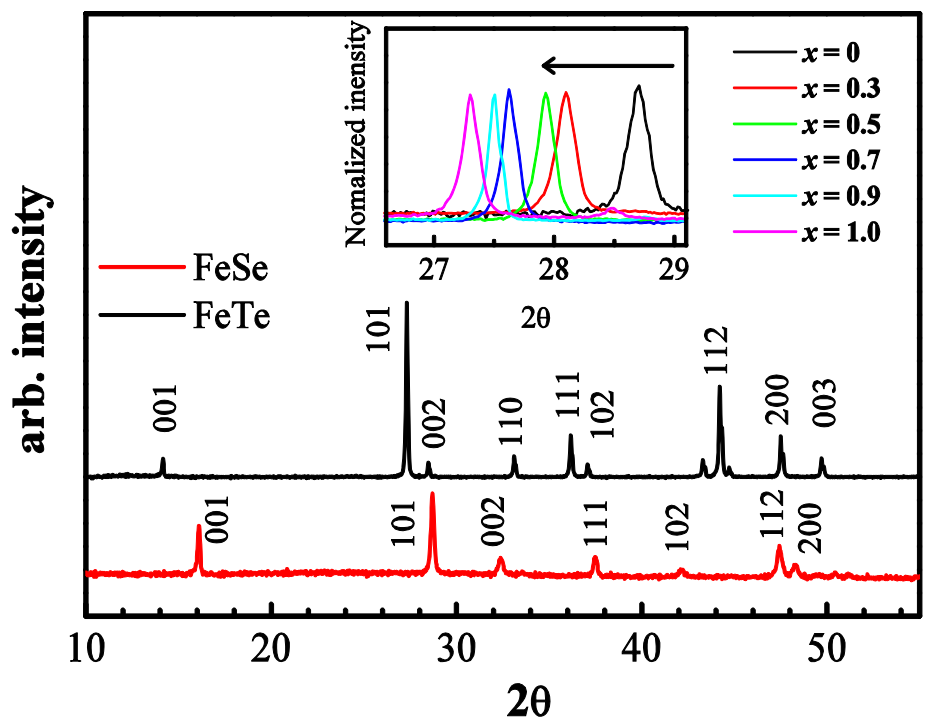

Figure 1. X-ray diffraction patterns of FeSe and FeTe using Cu-Kα radiation. Miller indices for tetragonal-PbO structure are shown. The (101) diffraction peak of a series of FeSe$_{1-x}$Te$_x$ concentrations is plotted in the inset. The arrow shows the direction of higher Te concentration.

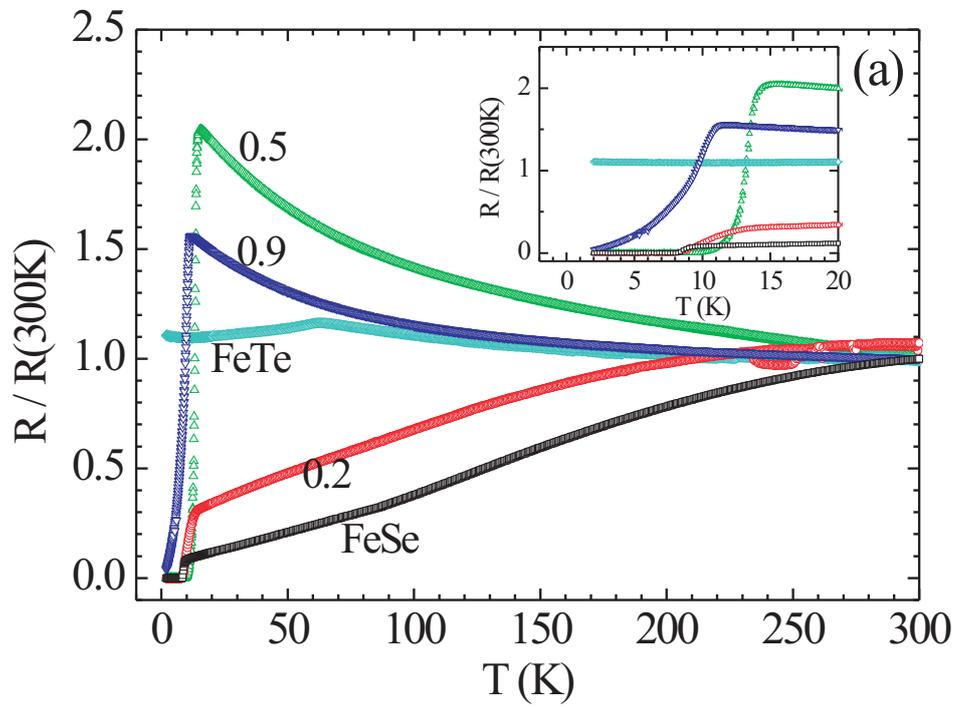

Figure 2. Temperature dependence of the normalized resistance with various Te concentrations at zero magnetic field. The inset displays the details near superconducting transition.

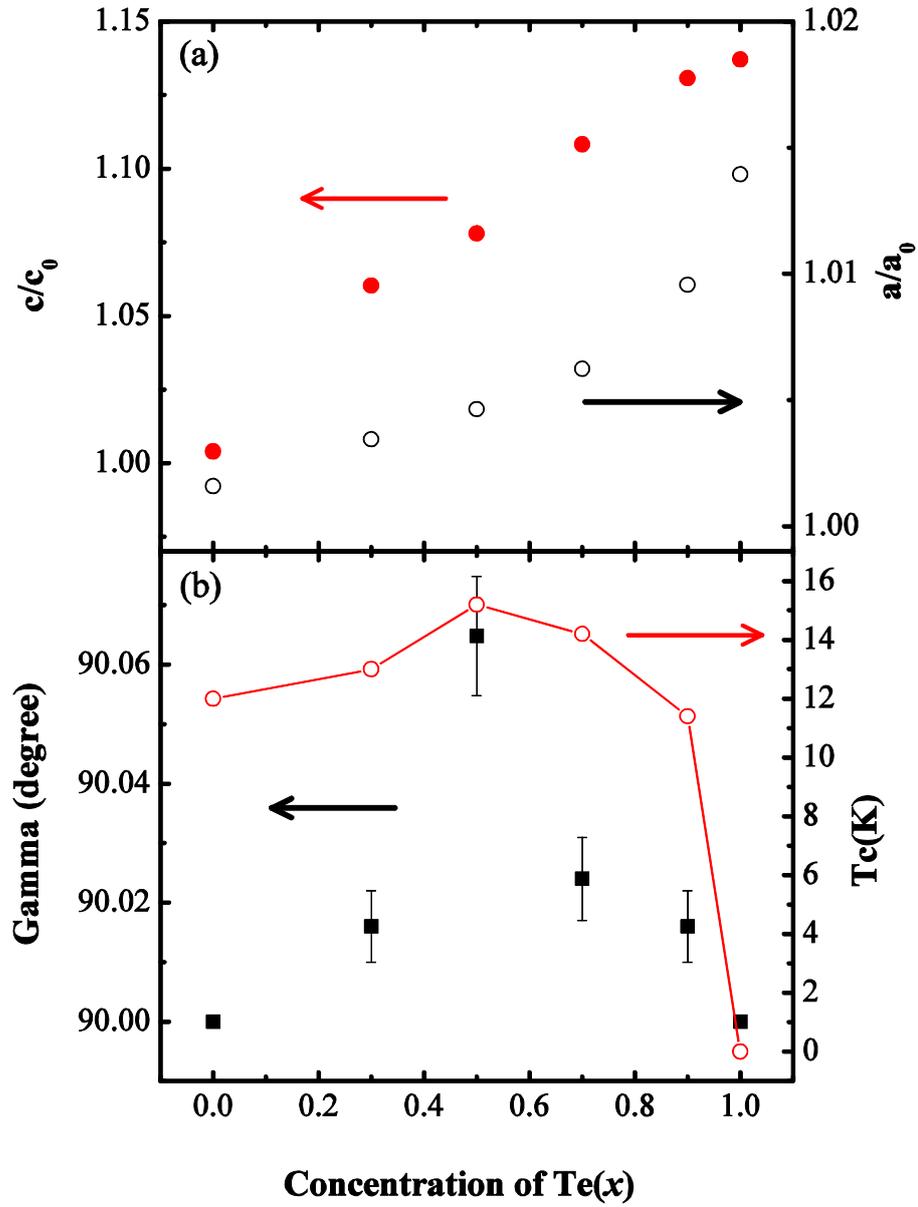

Figure 3. (a) Lattice expansion ratios of the tetragonal FeSe$_{1-x}$Te$_x$ obtained from X-ray refinement at room temperature.   $a_0$ and $c_0$ represent the lattice parameters of FeSe.   (b) The correlation between the gamma angle and transition temperature $T_c$ of various Te concentrations.   Here $T_c$ is defined as a temperature at which the resistance falls to 90 % of the normal state value.

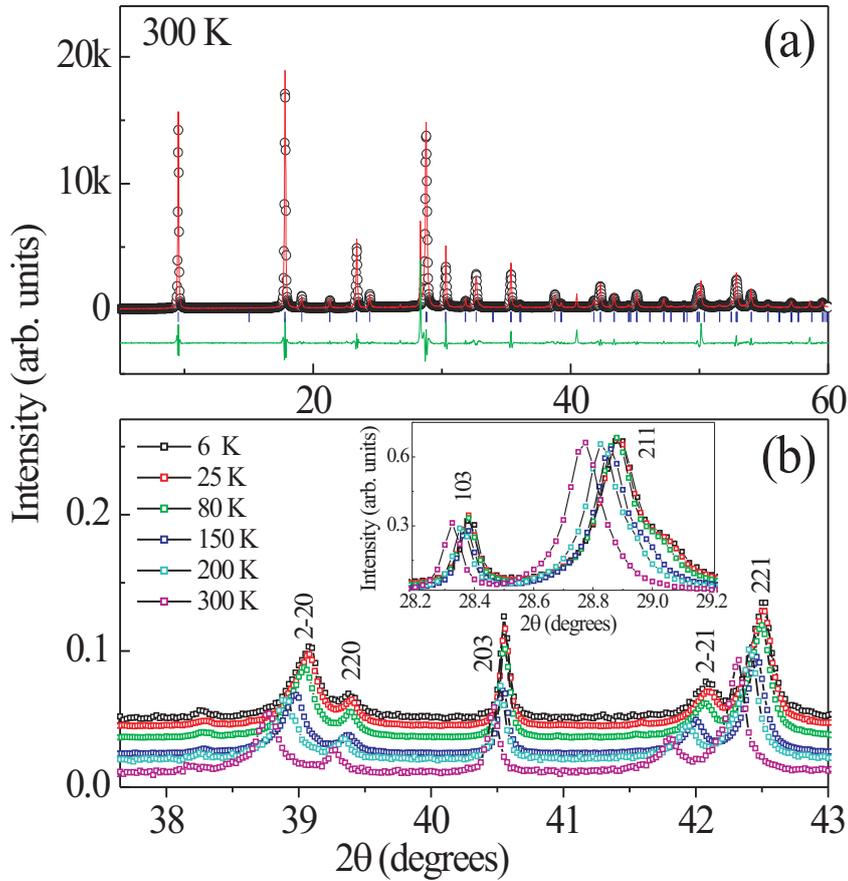

Figure 4. (a) The X-ray powder diffraction refinement for FeSe$_{0.5}$Te$_{0.5}$ at room temperature. Open black circles represent observed diffraction intensities. Red solid line shows calculated diffraction intensities using space group P-1, and green solid line is the result of subtracting the simulation results of tetragonal structure from the Bragg peak positions of the diffracted data. (b) Temperature dependence of the reflected Bragg peaks at (220), (203) and (221). The inset in (b) shows the (211) Bragg peak, which starts to split at around 100 K

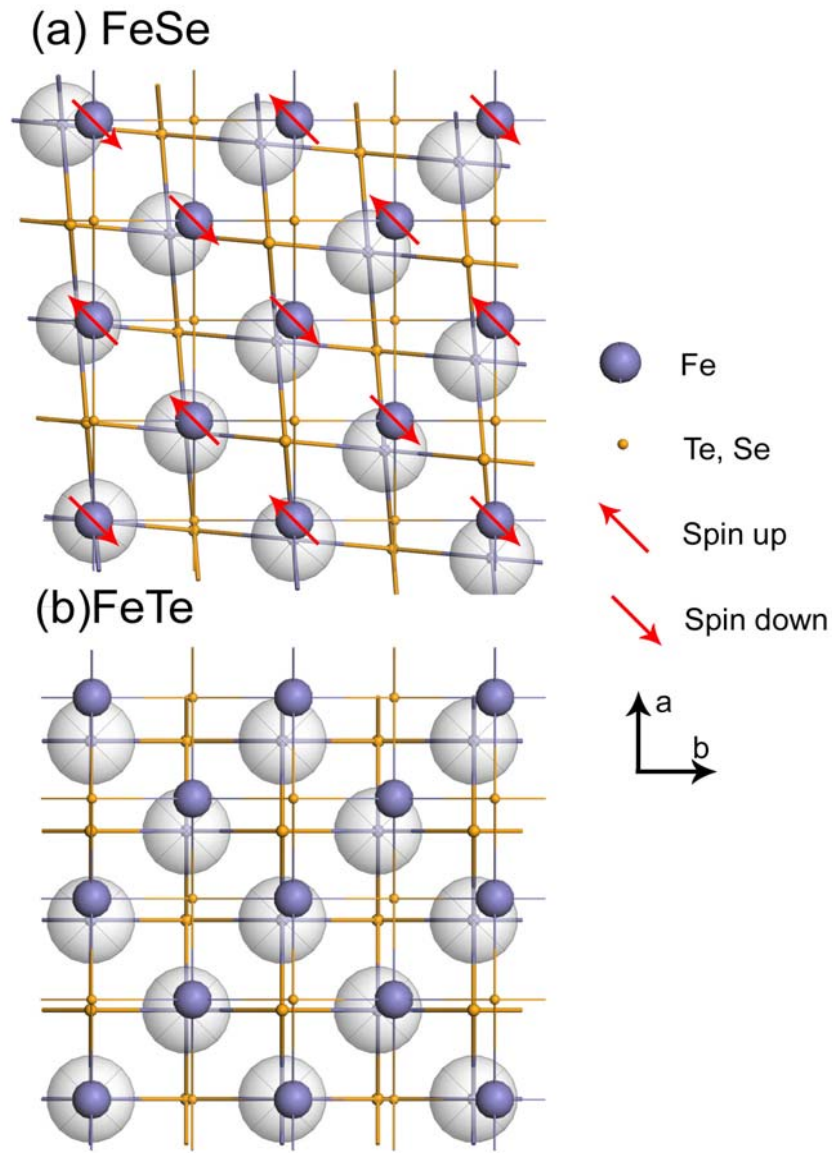

Figure 5. Schematic diagram of temperature dependence of (a) FeSe and (b) FeTe lattice. In this graph, blue spheres are Fe atoms, orange spheres are Se (or Te) atoms, and red arrows show the spin direction of Fe atoms. The matrix from larger to smaller sphere of Fe atoms schematically displays the room temperature to low temperature lattice distortion.